# Salty ice and the dilemma of ocean exoplanet habitability

Baptiste Journaux[1]

[1]Earth and Space Science Department, University of Washington, Seattle, WA, USA

**Habitability of exoplanet's deepest oceans could be limited by the presence of high-pressure ices at their base. New work demonstrates that efficient chemical transport within deep planetary ice mantles is possible through significant salt incorporation within the high-pressure ice.**

The search for and characterization of habitable liquid water extra-terrestrial environments is one of the most active areas of research in astrophysics, planetary science, and astrobiology[1–6]. The largest amount of liquid water in our universe by volume might be found in the gigantic oceans of water-rich exoplanets and icy moons. These vast ocean worlds might host some of the most promising environments for extra-terrestrial life to emerge and prosper. Nonetheless, the incredible pressures found within the water-rich layers will freeze the oceans from the bottom into dense high-pressure (HP) ices, separating the liquid layer from the nutrient-rich rocky interior (Fig. 1). Whether nutrients from the rocky core can still be delivered (or not) to the ocean through the ice to sustain life compatible conditions remains one of the outstanding dilemmas in modern astro- biology and planetary science[7]. With their new study, Hernandez et al.[8] show that HP ice layers inside ocean worlds can efficiently transport salts, further suggesting the high astrobiological potential of large water-rich worlds in our solar system and beyond.

## Distant and nearby ocean worlds and their frozen abysses

The growing catalog of exoplanets has revealed the widespread existence of a type of planet unknown in our solar system: ocean worlds. These planets, first suggested by ref. [7], are expected to possess a much higher amount of water than Earth. Despite its designation as the "blue marble", Earth is remarkably dry (<0.1 wt% $H_2O$) when compared to these worlds that may contain up to several 10 s wt% of $H_2O$. This would result in up to thousands of kilometers thick water-rich envelopes, called "hydrospheres", covering a rocky-metal core (Fig. 1). Depending on the size, two types of exoplanet bodies emerge: ocean super-Earths (>1.8·$R_{Earth}$) and mini-Neptunes (1.8–4·$R_{Earth}$ radius). Mini-Neptune's water-rich fluid envelopes are expected to be at several thousands of Kelvin, making them less relevant for astrobiology. Ocean exoplanets, such as the Trappist 1 e, f, g and h planets[9], are prime targets for habitability within their temperate oceans, possibly in equilibrium with their atmosphere, allowing for potential characterization and bio- signature detection by upcoming ground and space based observatories[1,4,7,10]. The icy worlds of the outer solar systems (Europa, Ganymede, Titan, and possibly Pluto), scaled-down versions of ocean exoplanets, also host large liquid water oceans below their frozen surfaces. Interest in the geo- physical and astrobiological attributes of these icy moons has motivated the most ambitious planetary exploration missions of the upcoming decade with NASA's Europa Clipper[11] and Dragonfly[12], and ESA's JUICE[13]. Furthermore, the search for biosignatures and habitable environments on exoplanets is one of the four main science goals of the newly launched James Webb Space Telescope[6].

As predicted for the largest icy moons (Ganymede, Titan and Calisto), the pressure within the hydrosphere of such planets can form dense HP water ices (ice II, III, V, VI, VII and X), stable below the liquid ocean[6] (Fig. 1). The presence of this HP ice layer separating the rocky core from the ocean is sometimes presumed to block geochemical exchange processes with the ocean. The resulting lack of chemical cycling and nutrient delivery from the rocky mantle to the oceans would limit the habitability potential of the ocean in such exoplanets and moons. Therein lies one of the most fascinating dilemmas of planetary habitability: could the exoplanets and icy moons with the most liquid water by volume be unable to sustain habitable condition?

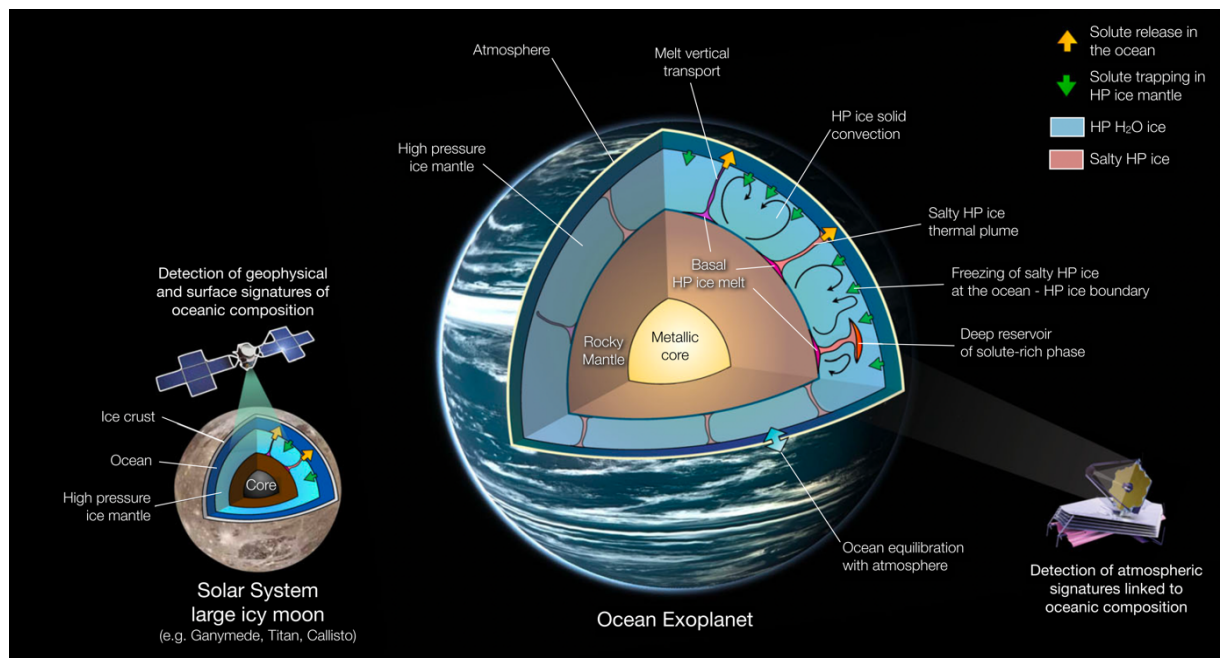

**Fig. 1** Schematic interior structure, dynamic and solute transport and cycling processes inside an ocean super-Earth and an icy moon of our solar system with high-pressure ice (Titan, Ganymede and Callisto). Basal melt enriched in solutes and nutrients at the base of the high-pressure ice mantle can be generated by various geological processes like hydrothermal sources, mineral aqueous alteration or volcanic activity. Vertical transport processes like buoyant liquid migration, salty high-pressure (HP) ice thermal plumes, solid-state convection and freezing at the bottom of the ocean into salty HP ice are reported. Possible trapping mechanisms like the formation of stable solute-rich phase reservoir within the HP ice mantle are also illustrated. The thermodynamic and chemical equilibration of the ocean with the atmosphere could lead to atmospheric spectral signatures linked to the ocean composition, detectable by modern and next generation telescopes like the JWST. For icy moons, geophysical and surface signature of the ocean composition will be investigated by upcoming planetary robotic mission ESA JUICE, NASA DragonFly. Background of icy moon Ganymede (NASA/Galileo); Ocean Exoplanet texture generated with PlanetMaker (http://planetmaker.apoapsys.com/).

This assumption is actively challenged by experimental and theoretical studies of icy mantle dynamics and HP ices chemical physics. Both suggest efficient transport of solutes from the rocky core to the upper ocean may be possible. First, dynamically, it has been demonstrated that efficient convection can happen within HP ice mantles in large icy moons like Titan and Ganymede, involving transport of liquid pockets from the hydrosphere-rock boundary to the ocean[14–16]. This remains to be studied for thicker exoplanet hydrospheres, but efficient liquid formation at the base of the HP mantle seems to be the rule rather than the exception[14,17]. Secondly, chemical physics advocates for efficient chemical transport through HP ices mantle.

## The discovery and mysteries of salty high-pressure ices

To understand the importance of this new study by Hernandez et al., one needs to re-examine a pillar of undergraduate physical chemistry teachings: "ice grows chemically pure from a saline solution by expelling all the salt ions into the liquid". This is generally true for low pressure common hexagonal ice (ice Ih). However, one remarkable discovery of ice physics of the last decade is that HP water ice with cubic crystalline structure above 20,000 atmospheres (phase VII and X) can incorporate significant amounts of simple ionic salts (e.g., NaCl, LiCl, or RbI salts) up to several weight %, while maintaining the same a crystalline structure[18–20]. These new type of crystalline solids in salt-water systems are also called "solid solutions" as the amount of salt is not stoichiometrically fixed, like with common types of salt hydrates such as hydrohalite (NaCl•2H$_2$O).

Several groups have reported intriguing properties of these salt- bearing ices. Salt incorporation distorts the HP ice crystalline lattice by breaking some of the hydrogen bound network[19,21]. Further- more the presence of ions in the lattice extend to higher pressure the stability of ice VII (molecular ice with H$_2$O molecule linked by H-bonding) at the expense of ice X (higher-pressure ionic ice with hydrogens in an symmetric position between oxygens)[22,23]. Con- troversy remained regarding the maximum amount of salt soluble[20,24], as well as the mechanism of salt ions inclusion being either interstitial (between H$_2$O molecules) or substitutional (replacing H$_2$O molecules). Furthermore, there is still uncertainty about solute incorporation effects on the lattice volume and density of salty ice. Experimental work by ref. [18] reported a reduction of the lattice volume, however, other recent experimental and theoretical work reported an increase[19–22]. How much salt can be incorpo- rated and how it could either reduce or increase the volume of HP ice (and affect its density), has crucial implications for our ability to describe the interior structure and dynamics of ocean exoplanets. Too high a density increase due to high salt solubility, or possible immiscibility within the HP mantle would inhibit convection and solute vertical transport[20]. Finally, limited data exists above 300 K on incorporation into HP ices, conditions found in exoplanets. To determine if solute transport is possible through planetary HP ice mantle, one needs to know, over the relevant range of pressure and temperatures, the solute solubility and its effect on density. Until now, we were lacking such a unified physical model and thermo- dynamic representation.

Hernandez et al.[8] provide a comprehensive study of NaCl incorporation in HP ice using state-of-the-art molecular dynamics simulations (first principle DFT-MD) over a wide range of condi- tions. By simulating crystallization of salty solutions, this extensive study binds together within a self-consistent physical model, several observations from previous studies more limited in scope and at lower temperatures. They report up to 2.5 wt% NaCl incorporation on substitutional sites that results in an increased lattice volume of salty HP ice. The study also provides a thermodynamic repre- sentation (equation of state) for salty HP ices that allows the authors to show that convective chemical transport through the dense salty HP ices is possible. Hernandez et al.[8] demonstrate that the density increase due to NaCl incorporation does not hinder vertical transport through solid-state convection. Along with previous stu- dies on convection of HP ice and melt vertical transport in mantles in smaller icy moons[14–16], this substantiates the notion that HP ice mantles do not act as chemical barriers between the rocky core and the liquid water ocean. The authors also suggest solute recycling mechanism with diffuse return flow of salty ice across the icy mantle-ocean boundary due to crystallization of salty ice (Fig. 1), further underscoring the permeability of HP ice mantles.

## Perspectives, near and far

Several fascinating questions remain open about salty ices to fur- ther constrain the potential habitability of extra-terrestrial oceanic environments on top of HP ice mantle. First, if simple salts like NaCl, LiCl, or RbI seem to be incorporated in HP ices in a similar manner, this remains to be quantified for larger solutes with molecular ions, relevant for planetary science and astrobiology, such as sulfates ($SO_4^{2-}$), or phosphates ($PO_4^{3-}$). Secondly, it has been shown that salt solubility can decrease by an order of mag- nitude in lower HP ice, like ice VI[20] expected in thick hydrospheres[6], or at high temperatures[24]. Due to these transitions, solute-rich ice could release its solutes within the HP ice mantle, forming a deep reservoir (Fig. 1) of solute-rich phases (e.g., solid salts, hydrates, clathrates). This could affect how some solutes can travel from a deep reservoir within the hydrosphere, to eventually equilibrate with the atmosphere and possibly become detectable from afar (Fig. 1). Unfortunately, few data exists on these systems at the relevant conditions. The thermodynamic stability of water- solute solid phases at HP (aka aqueous petrology), has received small attention despite its crucial importance in understanding the largest liquid water oceans of our solar system and beyond.

The study by Hernandez et al.[8] offers the most convincing argument yet in resolving the dilemma of large planetary hydro- sphere habitability. The upcoming planetary missions to Gany- mede, Callisto and Titan (ESA JUICE and NASA Europa Clipper & DragonFly) will directly test some of these conclusions by looking at the geophysical and surface signatures of the ocean composition and estimate the permeability of HP ice mantles (Fig. 1). These missions will not only allow us to better understand the inner- workings of the hydrospheres of icy moons, but will be key to understand the largest oceans in our universe in water-rich exo-planets, their potential for habitability and their future character- ization by modern and next generation telescopes.